Carbon Ejection from a $SiO_2$/SiC(0001) Interface by Annealing in High-Purity Ar

Takuma Kobayashi[1)] and Tsunenobu Kimoto[1)]

[1]Department of Electronic Science and Engineering, Kyoto University, Katsura, Nishikyo, Kyoto 615-8510, Japan

We found that carbon-associated byproducts formed at the dry-oxidized $SiO_2$/SiC(0001) interface could be decomposed and be taken out to the $SiO_2$ side by high-purity Ar annealing. We evaluated the concentration of the ejected carbon atoms in the $SiO_2$ by secondary ion mass spectrometry, and discovered that it clearly depends on the condition of oxide formation (dry-oxidation, nitridation treatment, and phosphorus treatment). This work provides an indirect but unambiguous evidence for the carbon-byproducts existing at the $SiO_2$/SiC interfaces, and also indicates that the phosphorus treatment removes the carbon-byproducts, leading to significant reduction of interface defects.

Silicon carbide (SiC) has been regarded as a suitable material for sustainable power electronics owing to its superior physical properties such as high critical electric field and wide bandgap.[1,2] The key device for power conversion, SiC metal-oxide-semiconductor field effect transistors (MOSFETs) have, however, suffered from the low channel mobility due to the extremely high interface state density ($D_{it}$ > $10^{13}$ cm$^{-2}$eV$^{-1}$) of silicon dioxide (SiO$_2$)/SiC systems.[2-5]

It has been believed that carbon-related byproducts created at the interface during the oxidation are the candidate of the interface defects.[2,5-8] However, direct detection of the carbon byproducts possibly residing at (or near) the interface has not been straightforward. Structural analyses based on x-ray photon spectroscopy (XPS), electron energy loss spectroscopy (EELS), and so on, have been performed to clarify the microscopic picture of the SiO$_2$/SiC interface so far.[2] However, these results are not always consistent to each other. Studies based on high-resolution XPS,[9-11] for instance, indicate that the interface is abrupt, and there exist only a few monolayers of sub-oxides.[11] Regarding studies based on transmission electron microscopy (TEM)/EELS,[12,13] one group showed a carbon-rich layer near the interface,[12] while another group insists that the interface is abrupt and that the existence of a transition layer is hard to detect by EELS.[13] The result of high-resolution medium energy ion scattering (MEIS) also shows that the interface is fairly abrupt.[14] These contradictions arise from the essential difficulty of the interface analysis. In normal structural analyses, such as XPS or EELS, the detection limit of foreign atoms is 0.3–1% of the host atoms.[2] Conversely, the imperfection of 0.1% of the host atoms will cause a huge number of electronic defects, which affect the electronic characteristics of the system.

Furthermore, most of the structural analyses focus on a single (or particular) sample(s) and very few of those deal with various samples owning different defect densities. Post-oxidation annealing (POA) such as interface nitridation (annealing in nitric oxide, NO[15,16] or $N_2O$[17,18]) or $POCl_3$ annealing (annealing in a gas mixture of $POCl_3$, $O_2$, and $N_2$[19,20]) is effective in passivating the interface defects. The underlying physics of reduction of the interface states owing to the POA has not yet been clarified, and thus it is important to reveal how the interface structure is changed when the oxidation or POA conditions are varied.

In this work, we succeeded in detecting the existence of carbon-associated byproducts at the SiC(0001) MOS interface indirectly but unambiguously. We discovered that it is possible to decompose the carbon-byproducts and take them out to the $SiO_2$ side by thermal annealing alone. We also found that the density of the ejected carbon atoms by the annealing strongly depends on the condition of gate oxide formation (dry oxidation, NO annealing, and $POCl_3$ annealing). From the obtained results, the mechanisms of interface passivation by NO and $POCl_3$ are discussed.

The samples employed in this study were 4H-SiC(0001) MOS structures. The doping density of the n-type epilayer was about $1\times10^{16}$ cm$^{-3}$. After standard RCA cleaning, dry oxidation was carried out at 1300˚C for 30–40 min. Subsequent NO annealing (10% diluted in $N_2$) or $POCl_3$ annealing (in a gas mixture of POCl$_3$, O$_2$, and N$_2$) were performed at 1250°C for 70 min or at 1000°C for 10 min, respectively. After the POCl$_3$ annealing, N$_2$ annealing was carried out at 1000°C for 30 min. To evaluate the $D_{it}$ distribution of the MOS structures, MOS capacitors were fabricated with circular Al electrodes with a diameter of about 500 μm. The $D_{it}$ was evaluated by a high-low method from the $C$-$V$ characteristics of the MOS capacitors. Secondary ion mass spectrometry (SIMS) was performed to monitor the carbon or nitrogen profile in SiC/SiO$_2$ systems. The primary ion species and the energies of the ions were O$_2^+$, 8 keV for carbon detection and Cs$^+$, 2 keV for nitrogen detection. The conversion of the measured secondary C ion counts to the C atom concentration was accomplished using a relative sensitivity factor determined by measuring a reference C-implanted SiO$_2$ sample. To cause diffusion of carbon species from the interface, annealing in high-purity Ar with a very low partial pressure of O$_2$ ($p_{O2}$) was performed in this study. The Ar gas was supplied from an Ar cylinder ($p_{O2}$ < 2 ppm) and then purified to obtain very low $p_{O2}$ (< 100 ppt). Note that the low $p_{O2}$ is a very important factor because the carbon atoms ejected during the annealing are easily transformed into gas species (CO or CO$_2$) when they meet O$_2$ inside the SiO$_2$.

Figure 1 shows the measured 1 MHz and quasi-static $C$-$V$ characteristics for the MOS structures (as-oxidized, NO-annealed, and POCl$_3$-annealed) prepared in this study. Large frequency dispersion appears in the sample oxidized at 1300°C, which is suppressed to some extent by the NO annealing at 1250°C, and after the POCl$_3$ annealing at 1000°C, the frequency dispersion almost disappeared. From the flat-band voltage shift of the 1 MHz $C$-$V$, the effective fixed charge density is estimated to be $5 \times 10^{11}$ cm$^{-2}$ (negative) for the as-oxidized sample, $1 \times 10^{11}$ cm$^{-2}$ (negative) for the NO-annealed one, and $3 \times 10^{11}$ cm$^{-2}$ (positive) for the $POCl_3$-annealed one. Figure 2 shows the interface state density evaluated by a high(1 MHz)-low method for the MOS structures (as-oxidized, NO-annealed, and $POCl_3$-annealed). The $D_{it}$ is highest in the as-oxidized sample, then in the NO-annealed one, and lowest in the $POCl_3$-annealed one, which agrees with a previous report.[19]

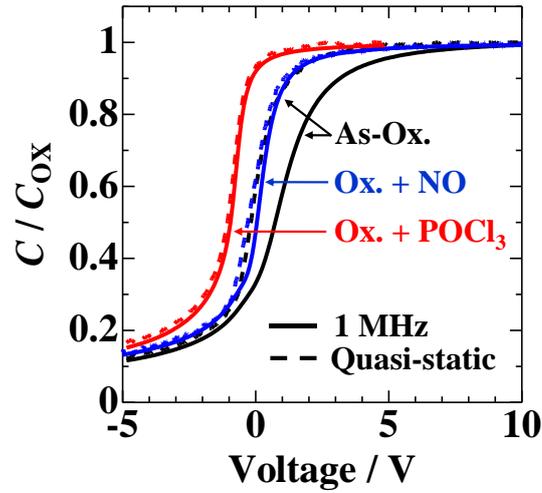

Fig.1: Measured 1 MHz and quasi-static $C$-$V$ characteristics for the SiC MOS structures (as-oxidized, NO-annealed, and $POCl_3$-annealed) prepared in this study. Dry oxidation, NO annealing, and $POCl_3$ annealing were carried out at 1300˚C, 1250˚C, and 1000˚C, respectively.

Then the MOS structures were annealed in high-purity Ar, and the carbon (and nitrogen) profile was evaluated by SIMS. Figure 3 shows the evaluated profile of carbon concentration for the as-oxidized (1300˚C) sample and that after Ar annealing at 1300˚C for 1 min. We can see that the carbon concentration in $SiO_2$ was close to the detection limit (~ $10^{18}$ cm$^{-3}$) in the as-oxidized sample. In contrast, after pure Ar annealing at 1300°C for 1 min, high concentration of carbon (> $10^{20}$ cm$^{-3}$) was detected inside the oxide. This result clearly indicates that the carbon-related byproducts exist at the as-oxidized MOS interface, and that the byproducts are decomposed and could be taken out to the $SiO_2$-side by thermal annealing alone. By integrating the carbon profile over the depth, we found that the carbon concentration in $SiO_2$ after the annealing is very high when converted into area density (~ $6\times10^{14}$ cm$^{-2}$).

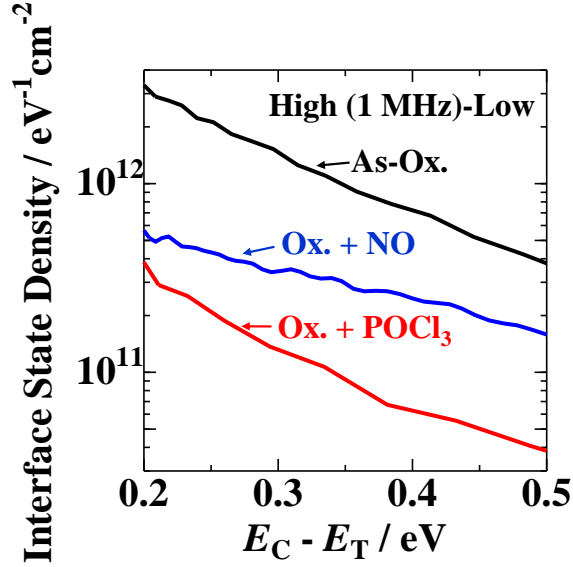

Fig.2: $D_{it}$ distributions for the SiC MOS structures evaluated by a high(1 MHz)-low method. Dry oxidation, NO annealing, and POCl₃ annealing were carried out at 1300°C, 1250°C, and 1000°C, respectively.

The post-oxidation annealing in NO is effective in both reducing the frequency dispersion of *C-V* curves and the interface state density, as shown in Figs.1 and 2. Then, we performed Ar annealing to the NO-annealed MOS structures, to investigate the mechanism of the observed defect passivation. The evaluated depth profiles of the carbon (and nitrogen) concentration are depicted in Fig.4. In the case of the NO-annealed sample, the carbon concentration in $SiO_2$ after Ar annealing at 1300˚C for 1 min, was close to the detection limit (~ $10^{18}$ cm$^{-3}$). After Ar annealing at 1300˚C for 5 min, however, high concentration of carbon is observed in the oxide (> $10^{20}$ cm$^{-3}$), which is comparable to or even a higher value than the case of the dry oxidized and Ar annealed (1300˚C, 1 min) sample (shown in Fig.3). This result suggests that the main effect of NO annealing is not the removal of the carbon-byproducts at the interface. In the nitrogen profile after the Ar annealing at 1300˚C for 5 min, we see that the high concentration of nitrogen ($\sim 10^{21}$ cm$^{-3}$) is localized near the interface, and this value is close to that just after the NO annealing.[21] In addition, the nitrogen concentration inside the oxide is around the detection limit ($\sim 10^{18}$ cm$^{-3}$) which is lower than the carbon concentration ($> 10^{20}$ cm$^{-3}$) by more than two orders of magnitude. From these results, it is likely that the nitrogen atoms are not directly attached to the carbon byproducts by the NO annealing.

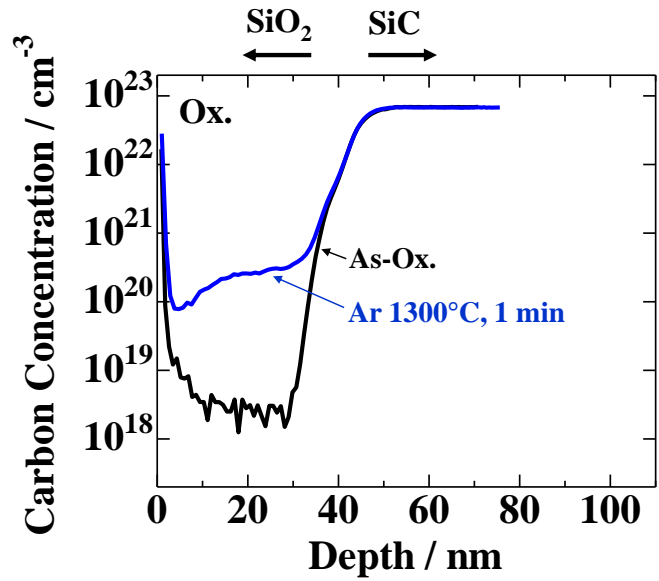

Fig.3: Depth profiles of carbon concentration in the dry oxidized and Ar-annealed SiC MOS structures evaluated by secondary ion mass spectrometry (SIMS). Dry oxidation and Ar annealing were both carried out at 1300˚C. Note that the profile for a few nm from the $SiO_2$ surface is unreliable due to the initial non-steady state in dynamic SIMS.

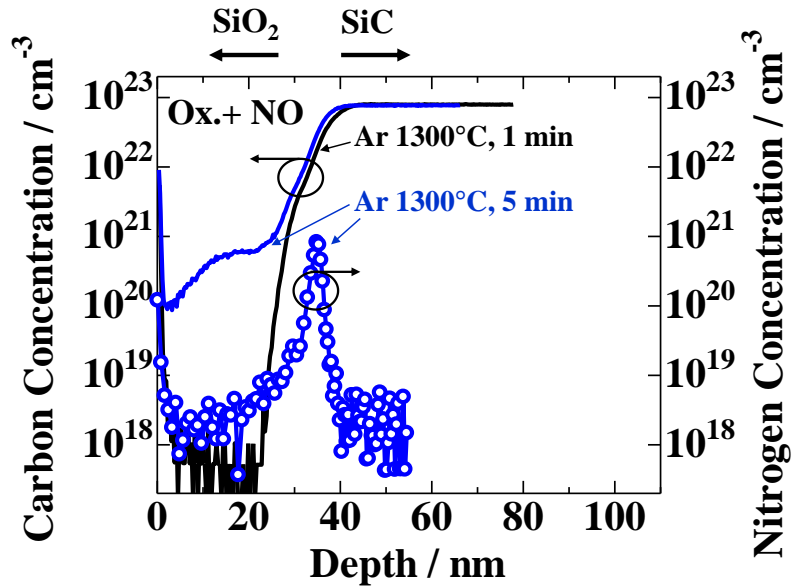

Fig.4: Depth profiles of carbon and nitrogen concentrations in the NO- and subsequently Ar-annealed SiC MOS structures evaluated by SIMS. Dry oxidation, NO annealing, and Ar annealing were carried out at 1300˚C, 1250˚C, and 1300˚C, respectively.

Regarding the origin of defect passivation owing to the NO annealing, it is suggested that the nitrogen atoms are doped in the SiC region near the interface and operate as donors, leading to formation of a very thin n$^+$ region in the SiC near the interface.[22)] Due to the formation of an n$^+$ region and resulting easier electron accumulation, surface band bending (or electric field) is overestimated in the $D_{it}$ analysis. This leads to underestimation of the energy difference between the surface Fermi level and the conduction band edge (and thus $E_C-E_T$ in Fig.2), which results in apparently lower defect densities. Such explanation does not conflict with our experimental facts. However, further investigations are required to clarify the mechanism of passivation owing to the nitridation process.

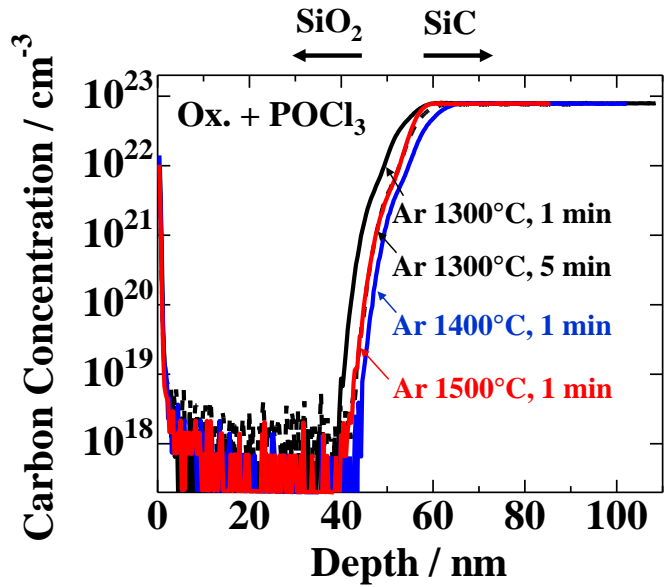

Fig.5: Depth profiles of carbon concentration in the POCl$_3$- and subsequently Ar-annealed SiC MOS structures evaluated by SIMS. Dry oxidation, POCl$_3$ annealing, and Ar annealing were carried out at 1300˚C, 1000˚C, and 1300–1500˚C, respectively.

As shown in Figs.1 and 2, POCl$_3$ annealing is more effective than NO annealing in passivating the interface defects. The POCl$_3$-annealed MOS structures were further annealed in Ar, to investigate the mechanism of the observed defect passivation. The depth profiles of the carbon concentration are shown in Fig.5. Unlike as-oxidized or NO-annealed samples, the carbon concentration detected in the oxide after the Ar annealing at 1300°C for 1–5 min was close to the detection limit (~ $10^{18}$ cm$^{-3}$). Furthermore, even after annealing at higher temperatures (1400°C and 1500°C), carbon was not detected. It should be noted that, in the case of as-oxidized or NO-annealed samples, the oxides completely sublimed by annealing at 1400°C for 1 min, and thus we speculate that the oxide sublimation process at such low oxygen pressure is related to destruction of the SiO$_2$ network due to the diffusion of carbon species from the interface. In the case of POCl$_3$-annealed sample, the diffusion of the carbon atoms does not occur, which keeps a solid network of the oxide, and thus the oxide survives even after Ar annealing at 1500°C.

From the above experimental results, it is highly likely that the majority of the carbon-associated byproducts induced during the oxidation process are removed by the subsequent POCl$_3$ annealing, leading to significant reduction of interface defects as shown in Fig.2. Such conclusion agrees with our recent prediction based on static and dynamic density-functional-theory (DFT) calculations.[23]

In summary, we compared the concentration of the ejected carbon atoms from as-oxidized, NO-annealed, and POCl$_3$-annealed SiC(0001) MOS interfaces by high-purity Ar annealing at 1300–1500°C. The carbon concentrations detected in the oxide after the annealing were very high (> $10^{20}$ cm$^{-3}$) in the case of as-oxidized or NO-annealed samples. The results not only indicate the existence of high concentration of carbon-associated byproducts at the as-oxidized MOS interface, but also suggest that the NO annealing is not removing the byproducts themselves. In contrast, the ejected carbon concentration was close to the detection limit ($\sim 10^{18}$ cm$^{-3}$) in the case of POCl$_3$-annealed sample. Thus, it is highly likely that the POCl$_3$ annealing is removing the carbon-related byproducts from the interface. This work provides an important data in revealing the microscopic picture of the SiC MOS interface and explaining the quality change of the interface owing to the post-oxidation annealing.

This work was supported in part by the Super Cluster Program from the Japan Science and Technology Agency.

1) B. J. Baliga, *IEEE Electron Device Lett.* **10**, 455 (1989).

2) T. Kimoto and J. A. Cooper, *Fundamentals of Silicon Carbide Technology* (John Wiley & Son Singapore, 2014).

3) N. S. Saks, S. S. Mani, A. K. Agarwal, *Appl. Phys. Lett.* **76**, 2250 (2000).

4) H. Yoshioka, J. Senzaki, A. Shimozato, Y. Tanaka, and H. Okumura, *AIP Advances* **5**, 017109 (2015).

5) T. Kobayashi, S. Nakazawa, T. Okuda, J. Suda, and T. Kimoto, *Appl. Phys. Lett.* **108**, 152108 (2016).

6) V. V. Afanas'ev, M. Bassler, G. Pensl, and M. Schulz, *Phys. Status Solidi A* **162**, 321 (1997).

7) R. H. Kikuchi and K. Koji, *Appl. Phys. Lett.* **105**, 032106 (2014).

8) Y. Matsushita and A. Oshiyama, arXiv:1612.00189 (2016).

9) B. Hornetz, H-J. Michel, and J. Halbritter, *J. Mater. Res.* **9**, 3088 (1994).

10) Y. Hijikata, H. Yaguchi, S. Yoshida, Y. Takata, K. Kobayashi, H. Nohira, and T. Hattori, *J. Appl. Phys.* **100**, 053710 (2006).

11) H. Watanabe, T. Hosoi, T. Kirino, Y. Kagei, Y. Uenishi, A. Chanthaphan, A. Yoshigoe, Y. Teraoka, and T. Shimura, *Appl. Phys. Lett.* **99**, 021907 (2011).

12) T. Zheleva, A. Lelis, G. Duscher, F. Liu, I. Levin, and M. Das, *Appl. Phys. Lett.* **93**, 022108 (2008).

13) T. Hatakeyama, H. Matsuhata, T. Suzuki, T. Shinohe, and H. Okumura, *Mater. Sci. Forum* **679-680**, 330 (2011).

14) X. Zhu, H. D. Lee, T. Feng, A. C. Ahyi, D. Mastrogiovanni, A. Wan, E. Garfunkel, J. R. Williams,

T. Gustafsson, and L. C. Feldman, *Appl. Phys. Lett.* **97**, 071908 (2010).

15) G. Y. Chung, C. C. Tin, J. R. Williams, K. McDonald, M. Di Ventra, S. T. Pantelides, L. C. Feldman, and R. A. Weller, *Appl. Phys. Lett.* **76**, 1713 (2000).

16) P. Jamet, S. Dimitrijev, and P. Tanner, *J. Appl. Phys.* **90**, 5058 (2001).

17) L. Lipkin, M. Das, G. Chung, J. Williams, N. Saks, and J. Palmour, *Mater. Sci. Forum* **383-393**, 985 (2002).

18) Y. Nanen, M. Kato, J. Suda, and T. Kimoto, *IEEE Trans. Electron Devices* **60**, 1260 (2013).

19) D. Okamoto, H. Yano, T. Hatayama, and T. Fuyuki, *Appl. Phys. Lett.* **96**, 203508 (2010).

20) T. Okuda, T. Kobayashi, T. Kimoto and J. Suda, *Appl. Phys. Express* **9**, 051301 (2016).

21) H. Yoshioka, T. Nakamura, and T. Kimoto, *J. Appl. Phys.* **112**, 024520 (2012).

22) G. Liu, A. C. Ahyi, Y. Xu, T. Isaacs-Smith, Y. K. Sharma, J. R. Williams, L. C. Feldman, and S. Dhar, *IEEE Electron Device Lett.* **34**, 181 (2013).

23) T. Kobayashi, Y. Matsushita, T. Okuda, T. Kimoto, and A. Oshiyama, arXiv:1703.08063 (2017).